\documentclass[preprint,12pt,sort,compress,oneside,onecolumn,a4paper]{elsarticle}

\usepackage[hidelinks]{hyperref}
\usepackage{url}

\usepackage{amssymb}
\usepackage{amsmath}
\usepackage{natbib}
\usepackage{booktabs}
\usepackage{tabularx}
\usepackage{siunitx}

\makeatletter
\def\ps@pprintTitle{%
 \def\@oddfoot{\em\small Preprint \hfill \today}%
 \def\@evenfoot{\em\small Preprint \hfill \today}%
}
\makeatother

\usepackage{graphicx}%
\usepackage{multirow}%
\usepackage{amsmath,amssymb,amsfonts}%
\usepackage{amsthm}%
\usepackage{mathrsfs}%
\usepackage[title]{appendix}%
\usepackage{xcolor}%
\usepackage{textcomp}%
\usepackage{manyfoot}%
\usepackage{booktabs}%
\usepackage{algorithm}%
\usepackage{algorithmicx}%
\usepackage{algpseudocode}%
\usepackage{listings}%
\usepackage{pdfpages}
\usepackage{bm}

\newcommand{\one}{^{\text{I}}}
\newcommand{\two}{^{\text{II}}}
\newcommand{\onetwo}{^{\text{I}\rightarrow\text{II}}}
\newcommand{\twoone}{^{\text{II}\rightarrow\text{I}}}
\newcommand{\oneortwo}{^{\text{I}/\text{II}}}
\newcommand{\twoorone}{^{\text{II}/\text{I}}}

\newcommand{\link}[1]{\href{#1}{\url{#1}}}

\begin{document}

\begin{frontmatter}



\title{Refining Machine Learning Potentials through Thermodynamic Theory of Phase Transitions}


\author[MFM]{Paul Fuchs} 
\author[MFM,MDSI]{Julija Zavadlav\fnref{cor1}}

\affiliation[MFM]{organization={Multiscale Modeling of Fluid Materials, Technical University of Munich},
            addressline={Boltzmannstraße 15}, 
            city={Garching},
            postcode={85748}, 
            state={Bavaria},
            country={Germany}}
\affiliation[MDSI]{organization={Atomistic Modeling Center, Munich Data Science Institute, Technical University of Munich},
            addressline={Walther-von-Dyck Straße 10}, 
            city={Garching},
            postcode={85748}, 
            state={Bavaria},
            country={Germany}}

\cortext[cor1]{Corresponding author (\href{mailto:julija.zavadlav@tum.de}{julija.zavadlav@tum.de})}

\begin{abstract}

Foundational Machine Learning Potentials can resolve the accuracy and transferability limitations of classical force fields.
They enable microscopic insights into material behavior through Molecular Dynamics simulations, which can crucially expedite material design and discovery.
However, insufficiently broad and systematically biased reference data affect the predictive quality of the learned models.
Often, these models exhibit significant deviations from experimentally observed phase transition temperatures, in the order of several hundred kelvins.
Thus, fine-tuning is necessary to achieve adequate accuracy in many practical problems.
This work proposes a fine-tuning strategy via top-down learning, directly correcting the wrongly predicted transition temperatures to match the experimental reference data.
Our approach leverages the Differentiable Trajectory Reweighting algorithm to minimize the free energy differences between phases at the experimental target pressures and temperatures. 
We demonstrate that our approach can accurately correct the phase diagram of pure Titanium in a pressure range of up to 5 GPa, matching the experimental reference within tenths of kelvins and improving the liquid-state diffusion constant.
Our approach is model-agnostic, applicable to multi-component systems with solid-solid and solid-liquid transitions, and compliant with top-down training on other experimental properties.
Therefore, our approach can serve as an essential step towards highly accurate application-specific and foundational machine learning potentials.

\end{abstract}



\end{frontmatter}

\section{Introduction}\label{sec1}

Molecular Dynamics (MD) is an important tool to discover new alloys for catalysis and manufacturing, understand their behavior, and optimize production processes.
For example, MD simulations provide microscopic insights into the thermal stability of high-entropy alloys~\cite{krounaAtomicScaleInsightsThermal2025} or how crystallization affects microstructure and mechanical properties~\cite{bizotMolecularDynamicsSimulations2023,mahat2019directionalSolidificationMD}.
Efficient high-throughput MD simulations enable active-learning guided search of high-temperature alloys~\cite{faracheActiveLearningMolecular2022}.
However, the typically employed classical potentials model interactions through simple functional forms with only a few tunable parameters.
Therefore, these potentials may fail to accurately reproduce multiple material properties with the same parametrization~\cite{delmastoInsightsCapabilitiesImprovement2024}.
More critically, parametrizations are not necessarily transferable across allotropes of the same element~\cite{castillo-sanchezTransferabilityClassicalPairwise2022}.
Thus, they can significantly fail to describe phase stability~\cite{castillo-sanchezTransferabilityClassicalPairwise2022,wenSpecialisingNeuralNetwork2021} and consequently the chemical potential important for crystallization~\cite{sunCrystallizationPhaseTransition2017}.

Machine Learning Potentials (MLPs) promise to overcome the limitations of classical force fields.
MLPs offer significantly improved accuracy, closely reproducing costly ab initio computations~\cite{zhangDeepPotentialMolecular2018,musaelianLearningLocalEquivariant2023,batatia2022mace}, while remaining scalable to million-atom systems~\cite{fuchsChemtrainDeployParallelScalable2025, musaelianLearningLocalEquivariant2023}.
Many MLPs have been tailored to specific metals and alloys~\cite{wenSpecialisingNeuralNetwork2021,zhaoGeneralpurposeNeuralNetwork2024} through a typical workflow of dataset generation, parameter optimization, and model validation~\cite{menonElectronsPhaseDiagrams2024}.
To cover a broad range of relevant system conformations, the datasets contain relaxed, stressed, and defective samples from multiple crystal structures, often generated with active learning and MD simulations at a wide range of temperatures~\cite{wenSpecialisingNeuralNetwork2021,zhaoGeneralpurposeNeuralNetwork2024}.
Still, when subsequently validated on phase stability predictions, the trained models only qualitatively reproduce experimental measurements but fail to match them closely within a magnitude of tens to hundreds of kelvins~\cite{wenSpecialisingNeuralNetwork2021,zhaoGeneralpurposeNeuralNetwork2024,menonElectronsPhaseDiagrams2024}.

Foundational MLP approaches aim to extend property prediction through MD simulation to a much wider class of chemically diverse materials by training on even more diverse and extensive datasets~\cite{batatia2025foundationmodelatomisticmaterials,wood2025umafamilyuniversalmodels,mazitovPETMADLightweightUniversal2025}. 
However, these models can exhibit significant deviations from experimental measurements~\cite{mannan2025evaluatinguniversalmachinelearning}.
For example, melting temperatures are sometimes underestimated by several hundreds of kelvins~\cite{mazitovPETMADLightweightUniversal2025}.
The reasons for the deviation of application-specific and foundation MLPs from experimental results are multifaceted.
Especially for foundational models, biases in the dataset towards specific structures or compounds can manifest as model bias~\cite{mannan2025evaluatinguniversalmachinelearning}.
Moreover, even carefully created and extensive datasets might omit important system conformations related to rare events~\cite{liuDiscrepanciesErrorEvaluation2023} or might suffer from systematic computational errors~\cite{kuryla2025accuratedftforcesunexpectedly}. 
Even if these errors are addressed through extensive benchmarks~\cite{mannan2025evaluatinguniversalmachinelearning,liuDiscrepanciesErrorEvaluation2023}, practically affordable density-functional theory calculations are often insufficiently accurate to describe experimental measurements of complex material properties~\cite{menonElectronsPhaseDiagrams2024,zhaoGeneralpurposeNeuralNetwork2024,imbalzanoModelingGaBinary2021}.
For example, different DFT functionals significantly over- or underestimate melting temperatures of many pure metals and semiconductors, sometimes in magnitudes up to hundreds of kelvins~\cite{dornerMeltingSiDensity2018,zhuEfficientApproachCompute2017,zhuPerformanceStandardExchangecorrelation2020}.
In principle, transfer learning strategies that extend the DFT dataset with more accurate electronic structure methods could further increase the accuracy of the MLPs~\cite{smithApproachingCoupledCluster2019}.
However, such methods significantly increase computational costs, which potentially render dataset generation, especially for foundational models or bulk systems, computationally intractable.

Instead of improving the underlying data quality, so-called top-down learning approaches ensure consistency with experiments by directly training MLPs to match experimental measurements~\cite{thalerLearningNeuralNetwork2021,rockenAccurateMachineLearning2024,rockenPredictingSolvationFree2024a}.
However, experimentally observable quantities are not direct outputs of the MLP, unlike force and energy, but are typically obtained by averaging system properties over long MD simulations.
Thus, top-down training requires more advanced algorithms, as directly applying automatic differentiation to such simulations does not guarantee meaningful gradients and might suffer from high memory requirements~\cite{thalerLearningNeuralNetwork2021}.
The model-agnostic DiffTRe~\cite{thalerLearningNeuralNetwork2021} algorithm leverages statistical mechanics reweighting approaches~\cite{zwanzigHighTemperatureEquationState1954,torrie1977} to efficiently compute gradients for equilibrium observables.
This computation remains efficient even when observables, such as fluctuation-based stiffness constants~\cite{thalerLearningNeuralNetwork2021,rockenAccurateMachineLearning2024} or free energies~\cite{rockenPredictingSolvationFree2024a}, require extensive sampling.
For only a few specified experimental data points, top-down learning of complex MLPs is ill-posed.
Purely top-down learned MLPs predict off-target properties with high uncertainty~\cite{thalerLearningNeuralNetwork2021}.
On the other hand, top-down training can be used to regularize or refine MLPs trained via Force Matching rather than for training MLPs from scratch.
This combination can correct pretrained MLPs for discrepancies with experimental data without inducing a large deviation from the underlying reference data~\cite{rockenAccurateMachineLearning2024}. 
Previous work has shown that MLPs can be refined top-down to predict solid-solid phase transitions at a target temperature~\cite{swinburne2025agnosticcalculationatomicfree}.
However, this approach is restricted to matching the free energy of linear models and is not directly applicable to solid-liquid phase transitions.

In this paper, we propose the model-agnostic {Differentiable Transition Temperature Correction (DiffTTC)} method to correct MLPs towards accurate predictions of phase diagrams.
The DiffTTC method adjusts the MLP using the model-agnostic and flexible DiffTRe algorithm such that the free energy difference between pairs of phases at target temperatures and pressures fulfills the thermodynamic criterion of phase coexistence.
As an example of titanium, we demonstrate that our algorithm can simultaneously correct for solid-solid and solid-liquid phase transitions.
Moreover, we assess that the procedure does not degrade the MLP performance on out-of-target properties and can be applied with other previously reported DiffTRe training targets~\cite{rockenAccurateMachineLearning2024}.
We evaluate our findings in the light of current foundational models, providing a path towards general-purpose MLPs that can predict complex material properties with high accuracy.

\section{Methods}

\subsection{Transition Temperature Computation}

We compute transition temperatures for the phase diagram using the thermodynamic theory of phase transitions and direct solid-liquid interface simulations. In thermodynamics, two phases I and II are in equilibrium if they have equal temperature $T$, pressure $P$, and chemical potential $\mu$~\cite{vegaDeterminationPhaseDiagrams2008}.
The chemical potential is directly related to the Gibbs free energy $G$ via $\mu = N G(P, T, N)$, which is a function of pressure, temperature, and number of particles $N$~\cite{tuckermanStatisticalMechanicsTheory2015}.
Therefore, Free Energy-based approaches determine the coexistence temperature $\hat T$ and pressure $\hat P$ by solving the equation
\begin{align}
    G\two(\hat T, \hat P) - G\one(\hat T, \hat P) = \Delta G\onetwo(\hat T, \hat P) = 0.\label{eq:coexistence_criterion}
\end{align}
These approaches typically employ two steps.
First, the approaches compute the free energy difference between the phases at a reference pressure or temperature, for example, using the Pseudo-Critical path method~\cite{correaRevisitingPseudosupercriticalPath2023} for solid-liquid transitions or the Frenkel-Ladd method~\cite{frenkelNewMonteCarlo1984} for solid-solid transitions (see next section).
Second, the approaches then calculate the change in free energy with varying pressure or temperature for both phases, for example, using multi-state reweighting~\cite{schieberUsingReweightingFree2018} or the Reversible Scaling method~\cite{dekoningOptimizedFreeEnergyEvaluation1999} (see next section).
In combination, these steps efficiently scan the free energy difference for a wide range of temperatures or pressures.

Free energy calculations require multiple complicated and costly simulations.
Especially the order-disorder transition in melting point calculations for (crystalline) solids is difficult to capture~\cite{correaRevisitingPseudosupercriticalPath2023}.
Therefore, alternative approaches have been proposed that dynamically estimate the stability of phases at different thermodynamic conditions~\cite{zhangComparisonMethodsMelting2012}.
High energy barriers might prevent phase transitions in simulation times accessible by MD and can cause effects such as supercooling and superheating.
Thus, direct approaches often introduce point defects into the solid phase or simulate systems with solid-liquid interfaces~\cite{zhangComparisonMethodsMelting2012}.
Point defects and interfaces both lower the energy barriers of melting and crystallization~\cite{zhangComparisonMethodsMelting2012}.
Simulating an interfacial system at constant pressure and energy but variable temperature additionally allows the system to equilibrate, possibly reaching the transition temperature at which both phases coexist in equilibrium~\cite{vegaDeterminationPhaseDiagrams2008}. 
To reduce the contribution of the interface to the total system pressure in an isoenthalpic-isobaric simulation, the system must be prepared such that the interface area is small compared to the dimensions of the box perpendicular to the interface~\cite{vegaDeterminationPhaseDiagrams2008}.
Then, the melting temperature at a given pressure can be found within a few trial simulations.

We perform the coexistence simulations in LAMMPS~\cite{thompsonLAMMPSFlexibleSimulation2022} using our plugin \texttt{chetrain-deploy}~\cite{fuchsChemtrainDeployParallelScalable2025}.
More details on the coexistence simulations are summarized in Supplementary Note 3.

\subsection{Non-equilibrium Free Energy Calculations}

Thermodynamic integration estimates the free energies between two potentials $u\one$ and $u\two$ by defining a $\lambda$-dependent potential $u(\lambda)$ with $\lambda \in [0,1],\ u(\lambda = 0) = u\one,\ u(\lambda=1) = u\two$ and computing the integral
\begin{align}
    \Delta F\onetwo = F\two - F\one = \int_{0}^1 \left\langle \frac{\partial u}{\partial \lambda}\right\rangle d\lambda = W_\text{rev}
\end{align}
equal to the reversible work $W_\text{rev}$~\cite{tuckermanStatisticalMechanicsTheory2015,dekoningOptimizingDrivingFunction2005}.
This integral can be approximated through numerical quadrature.
However, running multiple simulations to estimate the mean generalized force $\left\langle\frac{\partial u}{\partial \lambda}\right\rangle$ can be quite costly.
Therefore, non-equilibrium approaches~\cite{dekoningOptimizingDrivingFunction2005} approximate the reversible work by driving the system from $\lambda(0) = 0$ to $\lambda(t_s) = 1$ within a finite time $t_s$ to obtain the irreversible work $W_\text{irr}$ as an upper bound
\begin{align}
    W_\text{rev} \leq \int_0^{t_s} \dot \lambda (t) \frac{\partial H}{\partial \lambda(t)} dt = W_\text{irr}\onetwo,
\end{align}
where $H$ is the Hamiltonian, equal to the sum of kinetic and potential energy of the system. Similarly, driving the process backward provides a lower bound of the reversible work. If performed sufficiently slowly, both processes result in an unbiased estimator $W_\text{rev} = \left(\langle W_\text{irr}\onetwo\rangle - \langle W_\text{irr}\twoone\rangle\right)/ 2$~\cite{dekoningOptimizingDrivingFunction2005}.

The Frenkel-Ladd approach~\cite{frenkelNewMonteCarlo1984} is an efficient method for computing absolute free energies of crystalline solid phases and, consequently, the free energy difference between the phases. The approach constructs a potential transformation from the target potential to that of an Einstein Crystal, in which the $N$ particles with position $\mathbf r$ are harmonically coupled with a spring constant $k_E$ to their lattice position $\bm r_0$ via the potential $u(\lambda = 1) = \sum_{i=1}^N k_E(\mathbf r - \mathbf r_0)^2 / 2$. For this simple potential, the absolute free energy can be computed analytically. Thus, the sum of the absolute free energy of the Einstein and the free energy difference to the target potential results in the absolute free energy of the target crystal.
The Reversible Scaling path~\cite{dekoningOptimizedFreeEnergyEvaluation1999} provides the free energy change across multiple temperatures within a single simulation for each phase at constant pressure.
Thus, a single simulation for each phase is sufficient to find the coexistence temperature that solves equation~\eqref{eq:coexistence_criterion}.
Conceptually, the method maintains a constant simulation temperature $T\one$ but effectively scales the temperature to $T(\lambda) = T\one/\lambda$ by scaling the potential to $u(\lambda) = \lambda u\one$.
Non-equilibrium thermodynamic integration along $\lambda$ then results in the $\lambda$-dependent free energy $\Delta \tilde G^{\text{I}\rightarrow\lambda}$, which can be connected to temperature change in free energy via $\Delta G(T)= \Delta \tilde G^{\text{I}\rightarrow\lambda(T)}/\lambda + 3 Nk_BT\log\lambda/2$.

We perform the non-equilibrium simulations in LAMMPS~\cite{thompsonLAMMPSFlexibleSimulation2022} using our plugin \texttt{chetrain-deploy}~\cite{fuchsChemtrainDeployParallelScalable2025}, following the implementation from \citet{freitas2016}.
More details on the free energy integrations are summarized in Supplementary Note 4.

\subsection{Machine Learning Potential}
\label{subsec:mlp}

MLPs are a class of parametric models with parameters $\theta$ that derive forces $\bm f = -\frac{\partial u_\theta(\bm r)}{\partial \bm r}{}$ from a potential function $u_\theta(\bm r)$ that depends on all particle positions $\bm r$~\cite{unkeMachineLearningForce2021a}.
Often, the model architectures encode physical invariances, such as translational, rotational, and permutation invariance by design~\cite{unkeMachineLearningForce2021a}.
Descriptor-based potential, such as the DP model~\cite{zhangDeepPotentialMolecular2018}, ensures these invariances by encoding particle environments in invariant but often hand-crafted descriptors before feeding them particle-wise into general regression models such as fully connected neural networks.
More recent equivariant graph neural networks~\cite{batatiaDesignSpaceE3equivariant2025} learn these environment descriptors from data by performing message-passing between particles within a maximum cutoff distance.
Therefore, multiple message-passing operations propagate information beyond the cutoff, allowing the model to efficiently capture interactions between more distant particles.
Moreover, all operations are constructed such that the final descriptors remain invariant.
Therefore, these message-passing models provide the means to learn interatomic potentials purely from data.
Still, additive repulsive prior potentials have been shown to improve simulation stability, especially at high pressures~\cite{batatia2025foundationmodelatomisticmaterials}, and are thus often added to the model.

In this paper, we use an adapted version of the MACE~\cite{batatia2022mace} model.
MACE correlates local features to effectively construct high-order many-body messages. 
Therefore, only two message-passing iterations are sufficient to obtain a highly descriptive and efficiently parallelizable model~\cite{batatia2022mace}.
Unlike the original paper, we only use the output of the last, non-linear readout layer to learn the potential.
To improve the simulation stability, we combine the model with the pairwise Lennard-Jones potential 
\begin{align}
    u_\text{LJ}(\bm r) = \sum_{i=1}^N\sum_{j=1}^N 4\epsilon\left[\left(\frac{\sigma}{\lVert \bm r_i - \bm r_j\rVert}\right)^{12} - \left(\frac{\sigma}{\lVert \bm r_i - \bm r_j\rVert}\right)^{6}\right]
\end{align}
as an additive prior, where we infer the length scale $\sigma$ from the BCC lattice parameter $a=0.325\ \mathrm{nm}$ (see \citet{wenSpecialisingNeuralNetwork2021} for comparison) via $2 \sqrt[6]{2}\sigma=\sqrt 3a$ and choose $\epsilon=0.5\ \mathrm{kJ}\cdot\mathrm{mol}$.

We pretrain the MLP using the Force Matching method~\cite{ercolessiInteratomicPotentialsFirstPrinciples1994, fuchsChemtrainDeployParallelScalable2025} based on the curated dataset (see Supplementary information of \citet{rockenAccurateMachineLearning2024}) from \citet{wenSpecialisingNeuralNetwork2021}.
Therefore, we minimize the mean squared differences of the potential $u_\theta(\mathbf h_i, \tilde{\mathbf r}_i)$, force $\mathbf f_\theta(\mathbf h_i, \tilde{\mathbf r}_i) = -\frac{\partial u(\mathbf h_i, \tilde {\mathbf r}_i)}{\partial (\mathbf h_i\tilde{\mathbf r}_i)}$ and stress $\mathbf \sigma_\theta(\mathbf h_i, \tilde{\mathbf r}_i) = \frac{1}{\det \mathbf h_i}\frac{\partial u(\mathbf h_i, \tilde{\mathbf r_i})}{\partial \mathbf h_i}\cdot \mathbf h_i$~\cite{tuckermanStatisticalMechanicsTheory2015} to the DFT reference data $\hat u_i, \hat{\mathbf f}_i$, $\hat{\mathbf \sigma}_i$, predicted for the cell matrices $\mathbf h_i$ and fractional coordinates $\tilde{\mathbf r}_i$.
Thus, the loss function reads
\begin{align}
        \mathcal L(\theta) = \frac{1}{D}\sum_{i=1}^D\biggl[&\gamma_u\lVert u_\theta(\mathbf h_i, \tilde {\mathbf r}_i) - \hat u_i\rVert^2
        +\frac{\gamma_f}{3N_i}\lVert \mathbf f_\theta(\mathbf h_i, \tilde{\mathbf r}_i) - \hat{\mathbf f}_i\rVert^2\nonumber\\
        +w_i&\frac{\gamma_\sigma}{9}\lVert \sigma_\theta(\mathbf h_i, \tilde{\mathbf r}_i) - \hat{\mathbf \sigma}_i\rVert^2\biggr]\label{eq:fm_loss},
\end{align}
where $D$ denotes the number of training samples. The coefficients $\gamma_u$, $\gamma_{\mathbf f}$, and $\gamma_{\mathbf \sigma}$ balance the contribution of each quantity. 
We consider the virial only for uniformly strained samples by defining weights $w_i$ which are non-zero only for these specific samples and sum up to $\sum_{i=1}^D w_i = D$.
Details on the pretraining are summarized in Supplementary Note 1.

\subsection{Differentiable Transition Temperature Correction (DiffTTC)}
\label{subsec:diffttc}

The pretrained model, to which parameters we will refer to in the following as $\theta_0$, would correctly predict the transition temperature between phases if equation~\eqref{eq:coexistence_criterion} is fulfilled at the experimentally determined coexistence point.
In the DiffTTC method, we therefore adjust the model's parameters to correct the predicted transition temperatures by minimizing the magnitude of the predicted difference $\Delta G_{\theta}\onetwo(T_\text{exp}, P_\text{exp}) \stackrel{\theta}{\rightarrow} 0$ at the experimental coexistence point. 
To avoid first-order phase transitions and more complicated isobaric simulations, we adjust the Gibbs free energy by adjusting the Helmholtz free energy $F$.
If the refined MLP predicts the same molar volume $V\oneortwo/N$ as the pretrained MLP, the Gibbs and Helmholtz free energies change equally, i.e., $\Delta G\oneortwo_{\theta_0\rightarrow\theta } = \Delta F\oneortwo_{\theta_0\rightarrow\theta}$.
Hence, for each pair of phases $\text{I}$ and $\text{II}$, we aim to minimize the loss
\begin{align}
    \mathcal L(\theta) =\quad & \gamma_F \left(\Delta F\two_{\theta_0 \rightarrow \theta} - \Delta F\one_{\theta_0\rightarrow \theta} + \Delta G\onetwo_{\theta_0}\right)^2 \nonumber\\ + &\gamma_P  \left(P_\theta\one - P_\text{exp}\right)^2 + \gamma_P  \left(P_\theta\two - P_\text{exp}\right)^2,\label{eq:diffttc_loss}
\end{align}
where $P_\theta\oneortwo = \langle p_\theta\rangle\oneortwo$ defines the ensemble average $\langle \cdot \rangle_\theta$ of the instantaneous pressure $p_\theta$ over the canonical distribution $\rho_\theta(\bm r) \propto \exp(-\beta u_\theta(\bm r))$ at temperature $\beta=(k_BT)^{-1}$ ($k_B$ is the Boltzmann constant) and volume $V\oneortwo$.
The loss penalizes changes in the average pressure, where coefficients $\gamma_P$ and $\gamma_F$ control the penalty strength compared to the free energy target. The volumes $V\oneortwo$ and initial difference $\Delta G\onetwo_{\theta_0}$ are constants that have to be estimated only once prior to the DiffTTC method.

We optimize the loss using the Differentiable Trajectory Reweighting (DiffTRe) algorithm~\cite{thalerLearningNeuralNetwork2021}.
The DiffTRe algorithm expresses the ensemble averages of a quantity $a$ over the canonical distribution $\rho_\theta$ as weighted ensemble averages
\begin{align}
    \langle a(\bm r) \rangle_\theta = \langle w_\theta(\mathbf r)a(\mathbf r)\rangle_{\tilde \theta}
\end{align}
over the distribution $\rho(\bm r)_{\tilde \theta}$ with MLP reference parameters $\tilde \theta$, leading to the weights
\begin{align}
    \quad w_\theta(\mathbf r) = \frac{\exp\left(-\beta\left[u_\theta(\mathbf r) -u_{\tilde u}(\mathbf r)\right]\right)}{\left\langle \exp\left(-\beta\left[u_\theta(\mathbf r) -u_{\tilde u}(\mathbf r)\right]\right)\right\rangle_{\tilde \theta}}.
\end{align}
The algorithm then approximates the ensemble averages by drawing $S$ samples in an MD simulation and computing gradients through automatic differentiation.
Since only the weights and the instantaneous predictions, but not the sampled conformations, depend on $\theta$, the algorithm avoids propagating gradients through the simulation.
Moreover, DiffTRe reuses samples drawn in a previous simulation for multiple parameter updates if the effective sample size~\cite{carmichaelNewMultiscaleAlgorithm2012}
\begin{align}
    S_\text{eff} = \exp(-\sum_{i=1}^S w_i\log w_i),\label{eq:effective_sample_size}
\end{align}
measuring the statistical error of the approximation, remains above a predefined threshold $f < S_\text{eff}/S$.
If the effective sample size exceeds the defined threshold during the parameter update, DiffTRe updates the reference parameters $\tilde \theta \leftarrow \theta$ to the current parameters and draws new samples.
This resampling step limits the error in the predicted pressure, as resampling leads to equal weights, and fluctuation of the pressure samples depends only on the current potential. 
However, if the free energy is expressed as a simple ensemble average through the forward free energy perturbation estimator~\cite{zwanzigHighTemperatureEquationState1954}
\begin{align}
    \Delta F\oneortwo_{\theta_0 \rightarrow \theta} = -\frac{1}{\beta}\left\langle \exp\left(-\beta\left[u_\theta - u_{\theta_0}\right]\right)\right\rangle_{\theta_0},\label{eq:fep}
\end{align}
only a few samples with a large negative difference effectively contribute to the prediction if the difference between the current and reference potential becomes large.
Thus, the error in the predicted free energy difference can remain high even after resampling.
We circumvent this problem through the concept of thermodynamic integration as proposed in the ReSolv method~\cite{rockenPredictingSolvationFree2024a}.
The sequence of parameters $\theta[i]$, each connected to a trajectory, forms a thermodynamic integration path from the initial potential $\theta[0]=\theta_0$ to the current potential.
Thus, a more accurate differentiable estimator of the free energy can be obtained via the decomposition
\begin{align}
    \Delta F\oneortwo_{\theta_0\rightarrow\theta} = \Delta F\oneortwo_{\theta[n] \rightarrow \theta} + \sum_{i=0}^{n-1}\Delta F^{\oneortwo}_{\theta[i] \rightarrow \theta[i+1]}.
\end{align}
The free energy difference $\Delta F\oneortwo_{\theta[n] \rightarrow \theta}$ between the current parametrization $\theta$ and the most recent parametrization used to generate a trajectory $\theta[n]$ can be expressed via the forward-perturbation free energy estimator in equation~\eqref{eq:fep}.
The other differences $\Delta F^{\oneortwo}_{\theta[i] \rightarrow \theta[i+1]}$ can be computed immediately after each trajectory recomputation using, e.g., the more accurate Bennet Acceptance Ratio method~\cite{bennett1976}.

The ReSolv method and the DiffTRe method can train arbitrary MLPs to match pressures and target free energy differences.
However, our loss in equation~\eqref{eq:diffttc_loss} minimizes the free energy differences between two phases rather than fixed references.
Nevertheless, under first-order optimization of the loss, the gradients with respect to both phases $\mathrm{I}$ and $\mathrm{II}$ can be computed independently.
Therefore, we define the statepoint loss
\begin{align}
    \mathcal L\oneortwo(\theta) = \gamma_F\left(\Delta F\oneortwo_{\theta_0\rightarrow \theta} - \Delta \hat{F}\oneortwo\right)^2 + \gamma_P\left(P_\theta\oneortwo - P_\text{exp}\right)^2,\label{eq:statepoint_loss}
\end{align}
and update the free-energy targets before each optimization step to 
\begin{align}
    \Delta \hat{F}\oneortwo = \Delta F\twoorone_{\theta_0 \rightarrow \theta} \pm \Delta G_{\theta_0}\onetwo.\label{eq:free_energy_target}
\end{align}
When neglecting the gradients of $\Delta \hat{F}\oneortwo$, the sum of statepoint loss gradients $\nabla \mathcal L\one + \nabla\mathcal L\two = \nabla\mathcal L$ equals the gradients of the original loss function in equation~\eqref{eq:diffttc_loss}.

\begin{figure*}[tbh!]
    \centering
    \includegraphics[width=\linewidth]{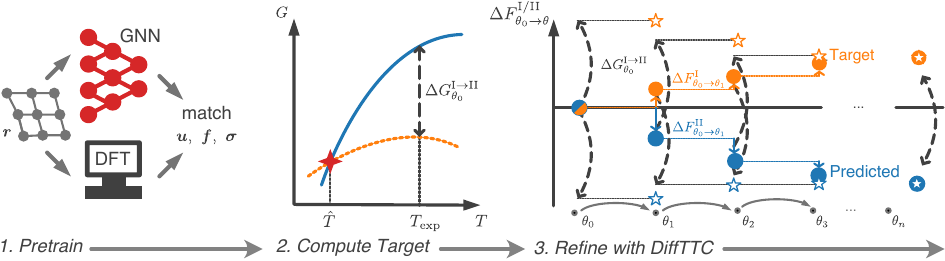}\vspace{-1.5em}
    \caption{\textbf{Differentiable Transition Temperature Correction (DiffTTC) Method.} 1. The method begins with a pre-trained potential. 2. Using the pretrained model with parameters $\theta_0$, the free energy difference $\Delta G\onetwo_{\theta_0}$ between phases I (blue line) and II (orange line) at the experimental transition temperature $T_\text{exp}$ is computed, e.g., by extrapolating the phase free energy change in temperature from the coexistence point (red star) at $\hat T$. 3. DiffTTC corrects the melting temperature by iteratively refining the potential parameter $\theta$, by matching the free energy changes $\Delta F_{\theta_0\rightarrow\theta}\oneortwo$ (orange and blue points) towards dynamically adjusted targets (orange and blue stars), eventually compensating the initial free energy difference (dashed arrow).}
    \label{fig:DiffTTC}
\end{figure*}

In summary, correcting the transition temperature with the DiffTCC method, as outlined in Figure~\ref{fig:DiffTTC}, consists of the following steps:
The procedure begins with a pre-trained MLP, obtained, for example, via Force Matching in equation~\eqref{eq:fm_loss} (Step 1).
The free energy differences between $K$ pairs of phases $\{(\mathrm{I},\mathrm{II})_i\}_{i=1}^K$ with average volumes $V\oneortwo_i$ at corresponding pressures $P_i$ and temperatures $T_i$ serve as training objectives. Thus, the quantities must be predicted beforehand (Step 2) but only once, as they remain constant during the optimization.
The DiffTRe algorithm iteratively updates the parameters to minimize the average loss in equation~\eqref{eq:diffttc_loss} for all statepoints (Step 3).
Therefore, the algorithm samples a batch of statepoint pairs and predicts the current free energy targets in equation~\eqref{eq:free_energy_target}.
If the effective sample size in equation $\eqref{eq:effective_sample_size}$ exceeds a predefined threshold of 0.9, similar to that in \citet{thalerLearningNeuralNetwork2021}, for any statepoint in the batch, the DiffTRe algorithm recomputes trajectories for all statepoints in the batch.
Therefore, the algorithm runs short simulations of the pure phases in the NVT ensemble. After recomputation or if the effective sample size is within the threshold, the algorithm computes the batch average statepoint loss, as given in equation~\eqref{eq:statepoint_loss}. Finally, the algorithm updates the potential parameters using a first-order optimizer, such as ADAM~\cite{kingmaAdamMethodStochastic2017}.

We use the DiffTTC method to refine the pre-trained MLP from Section~\ref{subsec:mlp}.
Therefore, we run the algorithm for 250 epochs with batches of $6$ pairs on a single A100 GPU.
More details on the optimization hyperparameters and convergence are given in Supplementary Note 2.

\section{Results}\label{sec2}

We benchmark our method by correcting a pretrained MLP as described in Section~\ref{subsec:mlp} using the DiffTTC method as described in Section~\ref{subsec:diffttc} based on experimentally determined phase stability data for pure Titanium (see \citet{wenSpecialisingNeuralNetwork2021}).

\subsection{Force and Energy Predictions}

First, we evaluate how the refinement affects the predicted forces and energies of the MLP in comparison to the DFT reference data, summarized in Table~\ref{tab:dft_errors}. The pretrained model matches the DFT reference with an energy error within a magnitude of several $\mathrm{meV}$ and force error within a magnitude of tens $\si{\milli\electronvolt\per\angstrom}$.
These error magnitudes are commonly reported for MLPs~\cite{liuDiscrepanciesErrorEvaluation2023,kuryla2025accuratedftforcesunexpectedly}. More specifically, the energy and force errors of our pre-trained model are comparable to those of the MLP presented by \citet{rockenAccurateMachineLearning2024}, which was trained on the same dataset.
Thus, the pretrained model serves as a reasonable baseline for the DiffTTC method.

\begin{table}[tbh]
    \centering
    \caption{Mean absolute errors of energies, forces, and virial on the test split of the curated~\cite{rockenAccurateMachineLearning2024} DFT dataset from \citet{wenSpecialisingNeuralNetwork2021}.}
    \label{tab:dft_errors}\vspace{1em}\footnotesize
    \begin{tabular}{lrr}\toprule
        \bf Quantity                                  & \bf Pretrained & \bf Refined      \\\midrule
        Energy \hfill[meV atom\textsuperscript{-1}]   &          3.9   &             21.7 \\
        Force \hfill[meV A\textsuperscript{-1}]       &         69.0   &             84.1 \\
        Virial \hfill [meV atom\textsuperscript{-1}]  &         26.3   &             37.8 \\\bottomrule
    \end{tabular}
\end{table}

For the DiffTTC-refined model, the energy, force, and virial errors are higher compared to the pretrained model, as expected. Notably, the energy errors are approximately half a magnitude higher than those of the pretrained model, while the force errors only increase by around $20\ \%$.
This difference can be due to the importance of energy in the long-term stability of phases. Slightly decreasing the energy for, e.g., the solid-state structures compared to liquid structures increases their relative probability.
If this change is approximately similar for all structures of a phase, the gradient of the energy mostly changes for rarely sampled transition states between the phases (see \citet{thalerDeepCoarsegrainedPotentials2022}, Figure~1, as an example illustration).

\begin{figure*}[tb!]
    \centering
    \includegraphics[width=\linewidth]{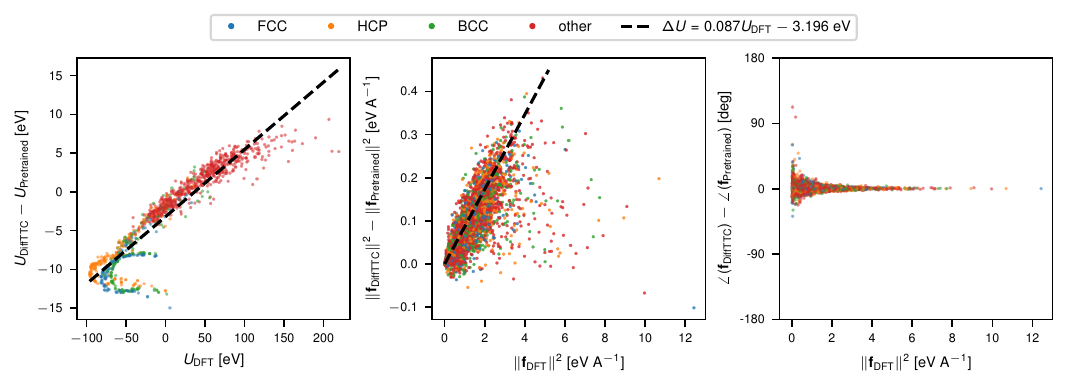}\vspace{-2em}
    \caption{\textbf{Predicted Energies and Forces:} Differences in the predicted energy and forces between the pretrained and the DiffTTC-refined model on the test split versus the DFT reference energy. The right plot shows the angle of predicted forces with respect to the DFT force. The black dashed line in the left plot corresponds to a linear regression fit to the energy differences. The color corresponds to the dominant crystal structure identified via the polyhedral template matching method~\cite{larsenRobustStructuralIdentification2016}. Only every 70th force component is shown.}
    \label{fig:force-energy-diff}
\end{figure*}

To assess the difference in force and energy predictions in more detail, we visualize them in Figure~\ref{fig:force-energy-diff} with respect to the DFT reference. The energy and force magnitudes are slightly higher, scaled by a factor around $1\ \%$. This scaling reinforces a higher energy for liquid structures compared to crystal structures and can decrease the relative probability of non-crystal structures. In addition to the scaling, lower energies are predicted for BCC structures, while higher energies are predicted for HCP structures. Thus, we expect that the stability of BCC structures is strengthened compared to HCP structures.
The direction of the predicted force changes only slightly, except for small force magnitudes, where even slight changes in the predicted forces can significantly alter the direction.

\subsection{Predicted Phase Diagram}

To assess how the DiffTTC training affects the stability of phases, we compute the phase diagrams predicted by the pretrained and refined models through solid-liquid coexistence simulations and non-equilibrium free energy integrations (see Supplementary Notes 1 and 2).
Figure~\ref{fig:phase-diagram} shows the predicted phase diagram for 6 pressures equally spaced between $0$ and $5\ \mathrm{GPa}$ in comparison to the reference data. Details on the error estimates are given in Supplementary Note 5.

\begin{figure*}[tb]
    \centering
    \includegraphics[width=\linewidth]{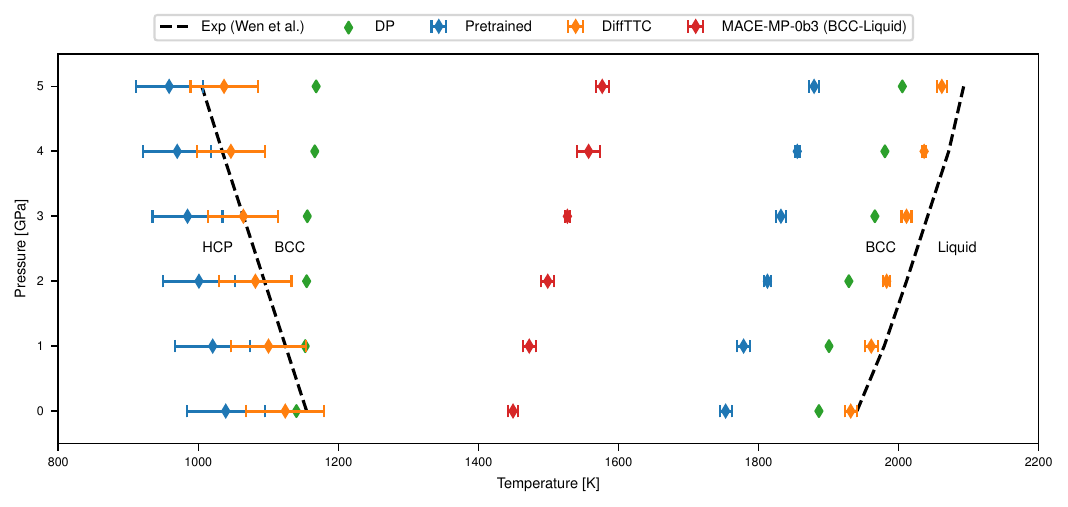}\vspace{-2em}
    \caption{\textbf{Phase-diagram:} Predictions of the pretrained and through DiffTTC refined MACE model are shown in comparison to the DP model~\cite{wenSpecialisingNeuralNetwork2021}, the foundational MACE-MP-0b3 model~\cite{batatia2025foundationmodelatomisticmaterials} (only BCC-liquid phase transition), and experimental measurements. The values for the DP model and the experimental data (black dashed lines) are extracted from~\citet{wenSpecialisingNeuralNetwork2021}. The error bars denote the uncertainty in the mean temperature in the coexistence simulations for the BCC-liquid transition and approximate the error in temperature of the free energy intersection point for the HCP-BCC transition based on error estimates reported in~\citet{freitas2016} for the free energy integration methods.}
    \label{fig:phase-diagram}
\end{figure*}

The phase diagram of the pretrained MLP aligns qualitatively roughly with the experimental reference.
Reference and predicted slopes of the HCP-BCC and BCC-Liquid phase boundaries closely agree.
According to the Clapeyron relation~\cite{vegaDeterminationPhaseDiagrams2008}, this agreement indicates that the model approximately captures the molar volume difference between the phases.
However, the absolute temperatures are predicted to be about $100$ to $150\ \mathrm{K}$ lower for the HCP-BCC transition and about $200\ \mathrm{K}$ lower for the BCC-liquid transition than the experimental data.

This discrepancy can derive from inaccuracies in the reference DFT data. Commonly used DFT functionals over- or underestimate melting temperatures by magnitudes sometimes up to 200 K~\cite{dornerMeltingSiDensity2018,zhuEfficientApproachCompute2017,zhuPerformanceStandardExchangecorrelation2020}. 
Optimally, with regard to the force matching objective, the MLP exactly reproduces DFT computations and therefore inherits these inconsistencies.
However, even if MLPs reproduce diverse DFT data with low errors, the MLP and DFT force and energy predictions might differ for rarely encountered samples, leading to deviations in predicted material properties~\cite{liuDiscrepanciesErrorEvaluation2023}.
To assess this variability, we compare our pretrained baseline to the DP model reported in \citet{wenSpecialisingNeuralNetwork2021}, which was selected and finetuned on extended DFT data from an ensemble of candidates based on its best overall test set performance (see \citet{rockenAccurateMachineLearning2024}, Supplementary Information, for a description of the differences in the data).
For the BCC-liquid phase boundary, the DP model trained on the same DFT dataset exhibits closer agreement to experimental measurements than the pretrained model, but still understimates the transition temperature.
On the other hand, the HCP-BCC phase boundary predicted by the DP model differs significantly more from the experimental reference than the boundary predicted by our pretrained and refined model.
The difference in the predicted phase diagram might derive from a higher preference for the energy error over the force error~\cite{goswamiHighTemperatureMelting2025}.
However, additional included structures and a different model architecture can also affect how the model interpolates unseen samples.
Nevertheless, the pretrained model and the DP model show pronounced differences from the experimental reference.
Thus, the results for the DP model emphasize that even for carefully designed project-specific datasets, additional refinement is necessary to achieve consistency with experiments.

We also test the foundational MACE-MP-03b model~\cite{batatia2025foundationmodelatomisticmaterials} to assess the out-of-the-box performance of models trained on large and diverse datasets.
Therefore, we calculate the BCC-liquid transition temperatures using the coexistence method as for our pretrained model, although with more than $50\ \%$ reduced interface area due to computational constraints.
Notably, the simulations still ran for around $12\ \mathrm{h}$, compared to less than $4\ \mathrm{h}$ for our simplified MACE models, on a single A100 (80 GB) GPU.
The foundational model predicts a melting temperature more than 400 K lower than the experimentally measured value.
This discrepancy is several hundred kelvins larger than the discrepancy between different experimental measurements~\cite{errandoneaSystematicsTransitionmetalMelting2001,stutzmannHighpressureMeltingCurve2015} (see Supplementary~Figure~S4 for a comparison).
Nevertheless, all simulations, even at high temperatures and pressures, have been stable and have led to a slope of the phase diagram similar to the experimental reference.
These observations agree with previous studies on the accuracy of thermodynamic properties predicted by foundational models. 
Although these MLPs show promising out-of-the-box performance on many computational benchmarks, they often cannot accurately capture complex material behavior~\cite{mannan2025evaluatinguniversalmachinelearning}.
Reasons for these failures include dataset biases towards specific compounds and structures~\cite{mannan2025evaluatinguniversalmachinelearning}.
Moreover, issues such as underrepresented important samples~\cite{liuDiscrepanciesErrorEvaluation2023}, which have been identified in application-specific datasets, are also likely to exist.
Thus, benchmarking these models on diverse predicted properties involving MD simulations or, even more, fine-tuning these properties to experimental data, as proposed in this work, can be crucial for obtaining adequately accurate foundational models.

Finally, we evaluate the phase diagram of the model refined by DiffTTC.
The DiffTTC-refined model predicts a phase diagram closely agreeing with the experimental reference, noticeably outperforming all other presented models.
For both transitions, the temperature is predicted within less than $50\ \mathrm{K}$ of the experimental reference.
For the HCP-BCC transition, the prediction agrees best at around $3\ \mathrm{GPa}$, while the temperature is slightly overestimated at higher pressures and slightly underestimated at lower pressures. For the BCC-liquid transition, the temperature is predicted best at $0\ \mathrm{GPa}$ and increasingly underestimated for higher pressures.
The slope of the predicted phase boundary remains slightly different from the experimentally determined slope but similar to the slope predicted by the pretrained model.
This similarity indicates that the predicted volumes for the phases are approximately preserved as intended.
Thus, correcting the predicted volume can be necessary to further improve the agreement with experimental data.
However, given accurate experimental references, such agreement can be ensured prior to the DiffTTC refinement using, e.g., the DiffTRe method~\cite{thalerLearningNeuralNetwork2021}.
Thus, incorporating the DiffTTC method into top-down training can tune the phase stability predicted for MLPs to align with experimental targets.

\subsection{Predicted Lattice Constants and Volumes}

\begin{figure*}[tb]
    \centering
    \includegraphics[width=\linewidth]{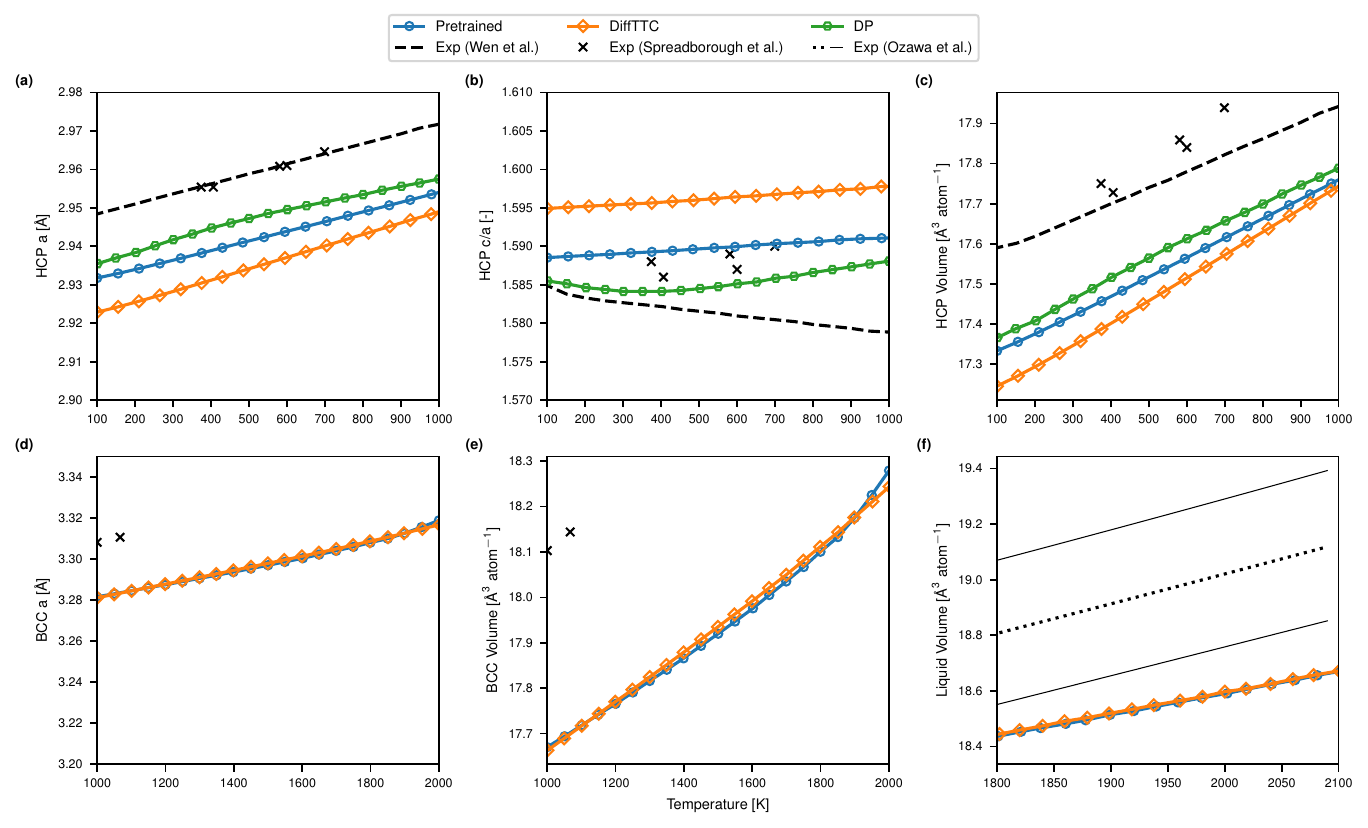}\vspace{-2em}
    \caption{\textbf{Lattice Parameters and Volumes:} Predictions of the pretrained and refined MACE model are shown in comparison to the DP model~\cite{wenSpecialisingNeuralNetwork2021} and experimental measurements. The values for the DP model and the experimental data corresponding to the black dashed lines in figures \textbf{(a-b)} are extracted from~\citet{wenSpecialisingNeuralNetwork2021}. The black crosses are taken from \citet{jspreadboroughMeasurementLatticeExpansions1959}.
    The regression line (dotted black) and uncertainty estimates (solid black) in plot \textbf{}{(f)} are converted from density measurements from~\citet{ozawaPreciseDensityMeasurement2017}.}
    \label{fig:volumes_and_lattice_constants}
\end{figure*}

The molar volumes affect the slope and thus the alignment between the predicted and experimental reference phase boundaries.
Moreover, we optimized the Helmholtz Free Energy in equation~\eqref{eq:diffttc_loss} instead of the Gibbs Free Energy based on the assumption that the pretrained and DiffTTC-refined MLPs predict similar volumes.
Therefore, we predict the volumes and lattice constants of the HCP, BCC, and liquid phases at multiple temperatures and zero pressure to assess whether our pressure penalty is effective. Figure~\ref{fig:volumes_and_lattice_constants} displays the predictions of the pretrained and DiffTTC-refined MLP in comparison to experimental measurements and values for the DP model reported in \citet{wenSpecialisingNeuralNetwork2021}.

The volumes predicted by the pretrained and DiffTTC-refined models closely agree over a wide temperature range for both the BCC and liquid phases, as well as for temperatures near the coexistence temperature for the HCP phase. Thus, the pressure penalty in the loss effectively preserves the predicted volumes at the coexistence temperature.
However, the predicted volumes for the HCP phase increasingly differ with lower temperatures. Moreover, the refined model predicts different HCP lattice constants than the pretrained model. These issues can arise from using the isotropic pressure in the pressure penalty. While the refined model predicts a similar average, resulting in a preserved volume, anisotropic stress leads to deformation of the box.
Thus, while the current pressure penalty is sufficiently effective to ensure an approximately equal change in Gibbs and Helmholtz free energy, an anisotropic pressure penalty could better preserve lattice constants.

Comparing the predicted volumes to experimental data, both the pretrained and refined models underestimate the volumes for all phases.
These deviations are specifically prominent at low temperatures, where quantum effects neglected by classical MD simulations become prominent~\cite{haoLatticeConstantsSemilocal2012}.
These effects can be modeled via Path-Integral MD simulations, but become insignificant at higher temperatures above $400-700\ \mathrm{K}$~\cite{yanAccuratePathintegralMolecular2022}.
For all temperatures, the HCP volumes predicted by the pretrained and DiffTTC-refined model are closer to the DP prediction than the DP prediction is to the experimental reference.
These deviations between the MLPs and experimental reference can be due to the accuracy of DFT calculations, which typically deviate from experimental lattice constants in a magnitude of tenths up to a few $\%$~\cite{haoLatticeConstantsSemilocal2012}. 
Thus, algorithms such as DiffTRe~\cite{thalerLearningNeuralNetwork2021} could improve the volumes predicted by the MLPs based on experimental data, thereby potentially aligning the slopes of the predicted phase boundaries.

\subsection{Out-of-Target Properties}

We evaluate structural and dynamic properties for the liquid state of titanium to assess how the DiffTTC method affects out-of-target properties.
Therefore, we compute the radial distribution function (RDF), angular distribution function (ADF), and diffusion constants displayed in Figure~\ref{fig:oot_props}. Details on the computation are given in Supplementary~Note~6.

\begin{figure*}[tb]
    \centering
    \includegraphics[width=\linewidth]{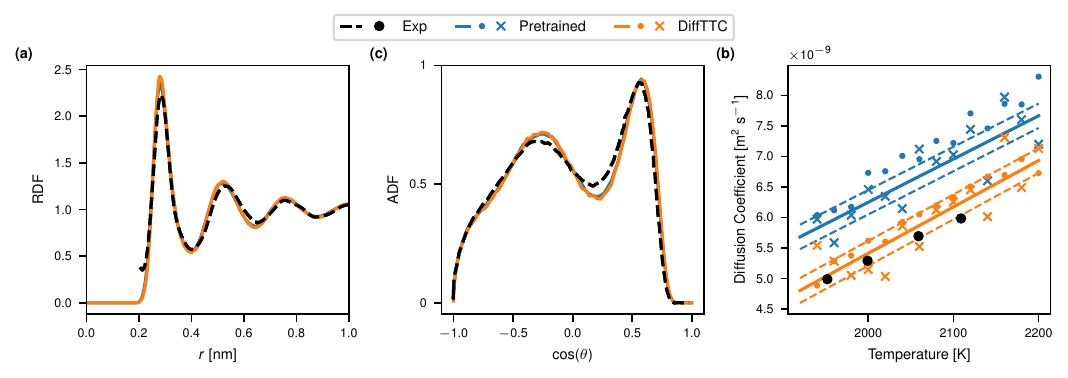}\vspace{-2em}
    \caption{\textbf{Out-of-Target Properties for Liquid Titanium.} \textbf{a} Radial Distribution Function (RDF) and Angular Distribution Function (ADF) predicted for the pretrained and DiffTTC potential at $0\ \mathrm{GPa}$ and $1960\ \mathrm{K}$ in comparison to experimental measurements for RDF~\cite{holland-moritzShortrangeOrderStable2007} and ADF~\cite{kimStructuralStudySupercooled2007}.
    \textbf{b} Diffusion constant at $0\ \mathrm{GPa}$ estimated for the pretrained and DiffTTC potential through velocity autocorrelation function (VACF, crosses) and temporal evolution of mean-squared displacement (points) in comparison to the experimental reference data~\cite{horbachImprovementComputerSimulation2009} (black points). The solid lines are a linear regression fit to the VACF measurements. The dashed lines denote two times the standard error of the regression line offset.}
    \label{fig:oot_props}
\end{figure*}

The RDF and ADF predicted by the pretrained and DiffTTC-refined MLP are equal to each other up to line thickness. This result is consistent with previous studies, that many off-target properties are not affected when top-down training highly descriptive MLPs~\cite{thalerLearningNeuralNetwork2021,rockenAccurateMachineLearning2024}. 
Moreover, the predicted RDFs and ADFs closely agree with experimental references. Although predictions do not match the experimental reference up to line thickness, the predicted peaks are closely aligned. Only the absolute values for the reference data are slightly lower than predicted by the MLPs.
The MLP-predicted diffusion constants differ more pronouncedly. The pretrained MLP slightly overstimates the diffusion compared to the experimental reference. On the other hand, the DiffTTC-refined model predicts a lower diffusion constant in good agreement with the reference
In the given example of self-diffusion, we can observe that refining the MLP through DiffTTC can have a positive impact on out-of-target dynamical properties.

\section{Discussion}

In this paper, we presented a method to correct an MLP to accurately reproduce experimentally determined phase stabilities.
Our method, DiffTTC, utilizes the DiffTRe algorithm to correct the free energy difference between two phases at the target pressure and temperature, ensuring that the thermodynamic criterion for the phases is fulfilled.
We tested our method on the example of pure Titanium, correcting the phase boundary of the HCP, BCC, and liquid phases in a pressure range from $0$ to $5\ \mathrm{GPa}$.

We demonstrated that our method can align the MLP-predicted phase diagram with an experimental reference.
Our presented DiffTTC-refined model significantly outperformed multiple reference models, including the pretrained baseline, the DP model from \citet{wenSpecialisingNeuralNetwork2021}, and the foundational MACE-MP-0b3~\cite{batatia2025foundationmodelatomisticmaterials} model.
The remaining minor deviations in the slope with respect to the experimentally determined phase boundary could potentially be resolved by correcting the model prior to the DiffTTC method to match experimentally determined lattice constants and molar volumes.
The DiffTTC method did not degrade the quality of the tested out-of-target properties. The RDF and ADF in the liquid phase predicted by the DiffTTC-refined model remained close to those predicted by the pretrained baseline model and in good agreement with experimental data.
Moreover, in the example of the diffusion constant, we could even observe a better agreement with experiments after correcting the phase boundaries.

Our method is independent of the procedure to compute the free energy difference between two phases for the initial MLP.
As long as the isolated phases remain stable at constant temperature and volume on the timescale of practical MD simulation, the DiffTTC method can refine the phases' free energy.
Thus, our method is directly applicable to more complex systems, such as binary or ternary phase transitions or phases with constant defect concentration.
Based on DiffTRe, our method is highly flexible in regard to computational efficiency and compatibility with other training approaches.
For example, simultaneously refining the MLP on mechanical properties~\cite{rockenAccurateMachineLearning2024} or structural properties~\cite{thalerLearningNeuralNetwork2021} is straightforward.
Moreover, enhanced sampling strategies such as population-based MD simulations~\cite{christiansenAcceleratingMolecularDynamics2019} could be used instead of largely sequential MD simulations for trajectory recomputations.

Our results strengthen the opinion that top-down training is a highly valuable addition to the common Force Matching training of MLPs~\cite{rockenAccurateMachineLearning2024, rockenPredictingSolvationFree2024a, mannan2025evaluatinguniversalmachinelearning}.
Problem-specific and foundational MLPs face multiple challenges in the dataset generation that often lead to a significant mismatch between predicted and experimentally determined material properties.
Our DiffTTC method is a promising approach to overcoming these issues through top-down learning due to its high flexibility.
In combination with a broad and accurate experimental dataset of molar volumes, as well as unary, binary, and potentially ternary phase diagrams, e.g., reliably collected in robotic laboratories, the DiffTTC method could highly improve the accuracy of foundational MLPs.
These models might then reliably predict phase diagrams that are difficult to access experimentally. 

In future work, we therefore aim to test the DiffTTC method for multi-component systems and train an accurate and transferable MLP. 
Therefore, we plan to test whether pretraining the MLP on volume data or matching the Gibbs Free Energy in the isobaric-isothermal ensemble might correct the slopes of the phase boundaries. 
Moreover, we plan to test whether data from the same phase across different temperatures and pressures can be shared or reused to reduce the computational cost associated with running molecular dynamics simulations independently for each temperature and pressure.

\section*{CRediT}

\textbf{Paul Fuchs:} Conceptualization, Methodology, Software, Formal analysis, Investigation, Visualization, Writing - Original Draft. 
\textbf{Julija Zavadlav:} Conceptualization, Validation, Writing - Review \& Editing, Resources, Funding acquisition, Supervision, Project Administration.

\section*{Acknowledgements}

This work was funded by the Deutsche Forschungsgemeinschaft (DFG, German Research Foundation) - 534045056.
The authors gratefully acknowledge the Gauss Centre for Supercomputing e.V. (www.gauss-centre.eu) for funding this project by providing computing time through the John von Neumann Institute for Computing (NIC) on the GCS Supercomputer JUWELS~\cite{juwels} at Jülich Supercomputing Centre (JSC).

\section*{Data and Code Availability}

The DFT Ti-dataset~\cite{wenSpecialisingNeuralNetwork2021} in curated form is publicly available at \link{https://github.com/tummfm/Fused-EXP-DFT-MLP}.
The experimental phase transition data were extracted from \citet{wenSpecialisingNeuralNetwork2021}.
The MD software LAMMPS and the training software \texttt{chemtrain} are publicly available at \link{https://github.com/lammps/lammps} and \link{https://github.com/tummfm/chemtrain}.
The training and evaluation scripts will be made publicly available at \link{https://github.com/tummfm/DiffTTC} upon acceptance of the manuscript.

\begin{appendix}
\end{appendix}

\bibliography{sn-bibliography}

\end{document}